\documentclass[traditabstract]{aa}
\usepackage{graphicx}
\usepackage{natbib}
\bibpunct{(}{)}{;}{a}{}{,}
\usepackage{txfonts}
\begin{document}
%
\title{ Detection of the ellipsoidal and the relativistic
 beaming effects in the CoRoT-3 lightcurve
}
\author{
     T.\ Mazeh           
\and S.\ Faigler         
\\
}
\institute{ 
School of Physics and Astronomy, 
Raymond and Beverly Sackler Faculty of Exact Sciences,\\ 
Tel Aviv University, Tel Aviv  69978, Israel
}
\date{Received / Accepted }

\abstract{ 
CoRoT-3b is a 22 Jupiter-mass massive-planet/brown-dwarf object,
orbiting an F3-star with a period of 4.3 days. We analyzed the
out-of-transit CoRoT-3 red-channel lightcurve obtained by the CoRoT mission 
and detected the ellipsoidal
modulation, with half
the orbital period and an amplitude of $59\pm9$ ppm (parts per million), and 
the relativistic beaming effect, with the orbital
period and an amplitude of $27\pm9$ ppm. 
Phases and amplitudes of both modulations are consistent with
our theoretical approximation. 
}
\authorrunning{Mazeh \& Faigler}  
\titlerunning{Ellipsoidal and beaming effect in CoRoT-3}	       
\maketitle

\section{Introduction}        
\label{introduction}                            

Close binary stellar systems display two well-known periodic
photometric modulations --- the ellipsoidal effect, due to the
distortion of each
component by the gravity of its companion \citep[see a review
by][]{mazeh08}, and the
reflection/heating effect (referred to here as the reflection effect), 
induced by the luminosity of each
star, which affects only the close side of its companion \citep[e.g.,][]{for10}.
These two effects can be observed even for non-eclipsing binaries, but
are much easier
to study in eclipsing binaries, where the binarity of the system and
the phases of the orbital motion are well known from the observations
of the eclipses. Most algorithms that analyze lightcurves of eclipsing
binaries, such as EBOP \citep{etzel80, popper81} 
and its derivative EBAS (Tamuz, Mazeh 
\& North 2008), WD \citep{wilson71}, and ELC \citep{orosz00},
include by default these two effects in their model of the
out-of-eclipse lightcurve.

A much smaller and less studied photometric modulation is the
relativistic beaming effect, sometimes also called Doppler boosting,
induced by the stellar motion relative to the observer --- $V_{rel}$,
whose amplitude is on the order of $V_{rel}/c$, where $c$ is the
velocity of light. Before the era of space photometry this effect has been
noticed only once, by Maxted, March \& North (2000), who observed
 KPD 1930+2752, a binary with a very short period, of
little longer
than 2 hours, and a radial-velocity amplitude of 350 km/s. The
beaming effect of that system, which should be on the order of $10^{-3}$, was
hardly seen in the photometric data.

Space photometry, which was developed to detect the minute transits of
exoplanets, has substantially improved the precision of the
produced lightcurves. The CoRoT \citep{rouan98, baglin06, auvergne09} and Kepler
\citep{borucki10, koch10} missions are producing  hundreds of
thousands of continuous photometric lightcurves with timespan of tens
and hundreds of days, with precision that can reach as high as
$10^{-3}$--$10^{-4}$ per measurement. 
It was therefore anticipated that CoRoT and Kepler
should detect all three effects \citep[e.g.,][]{drake03}, in particular
the beaming
effect for both planets \citep{loeb03} and eclipsing binaries (Zucker, Mazeh \& Alexander 2007).

As predicted, 
\citet{vankerkwijk10} detected in the Kepler lightcurve 
the ellipsoidal and the
beaming effect of two eclipsing binaries,
KOI 74 and KOI 81 (Rowe 2010). They used the
radial-velocity photometric beaming effect to derive the mass of the
secondary in the two systems and showed that in both cases it
was probably a white dwarf.
 \citet{welsh10} identified the ellipsoidal
effect in the Kepler data of HAT-P-7, a system with a known planet 
of 1.8
Jupiter masses (=$M_{Jup}$) and a period of 2.2 days \citep{pal08}. 
\citet{snellen09} detected in the CoRoT data the reflection effect of CoRoT-1.

In this paper, we report the detection of the ellipsoidal and the
beaming effects of CoRoT-3, induced by its
massive-planet/brown-dwarf companion. CoRoT-3b \citep{deleuil08} is a
22 Jupiter-mass object, orbiting an F3-star with a period of 4.26 days. 
The stellar rotation is 
probably synchronized with the orbital period.
We analyzed the CoRoT-3 out-of-transit
red-channel lightcurve and detected
two modulations, one with the orbital period and the other with its
first harmonic.
We attributed the two modulations to the beaming and the ellipsoidal
effects, respectively, as their phases and amplitudes were consistent 
with our order-of-magnitude approximation. 
Sect. 2 presents our data analysis, Sect. 3
compares our findings
with theoretical approximations, and Sect. 4 summarizes our results.

\section{Data analysis}        
\label{analysis}                            
CoRoT-3 was discovered \citep{deleuil08} in the data obtained during
the first long run of the CoRoT mission --- LRc01, which lasted for 152.012 d, from 
May 26 until October 25, 2007 
\citep[for details about this run see][]{cabrera09}. 
The optics of the mission include a bi-prism that disperses the stellar light into three channels, 
{\it red}, {\it green}, and {\it blue}, the sum of which is called the {\it white} channel. 
For bright stars, including CoRoT-3, 
the light intensity coming 
through each of 
the three channels is available. We used the so-called 
N2 data level \citep{baudin06} of CoRoT-3, which is now public. 

As the modulations we searched for were quite small, we had to
prepare and clean the data before searching for any periodic effect.
We decided to concentrate on the red-channel data, because this channel included 
most of the stellar light detected by CoRoT, and the other two channels only added
noise to the data (see below). \citet{snellen09} adopted a similar strategy 
when analyzing the lightcurve of CoRoT-1. 
This section describes how we 'cleaned' the data, removed the long-term variation, 
and searched for
the periodic modulation with the orbital period and its harmonics. 

\subsection{Cleaning the lightcurve}

The 'cleaning' of the CoRoT-3 lightcurve had the following stages:
 
\begin{itemize}
\item 
Rebinning: Corot-3b was detected before the CoRoT run was completed 
and therefore the
cadence of the
observations was changed during the run --- the first part of the
lightcurve is composed of 512 s exposures, while the later part
contains 32 s exposures. Since we were interested in modulations with
periods equal to or longer than 
half
the orbital period, at about 2.1 d, the entire light curve was
re-binned into 512 s bins. Altogether, we derived 22,072 valid measurements.

\item
Removing transits: 1165 measurements taken during the transits 
of CoRoT-3 were removed
from the analysis.

\item
Jump removal: One 'jump', at CoRoT HJD of 2746.99, 
probably caused by a 'hot pixel' event, was identified and corrected. The counts after
the jump were adopted to the
stellar flux before the jump, while 24 measurements following the jump
were removed.
\item
Outlier removal: We identified 106 outliers by 
calculating the running median and RMS around each point, and rejecting
measurements that differed by $4\sigma$ or more from their
corresponding median. We were left with 20,801 data points.

\end{itemize}

\subsection{Long-term detrending with a cosine filter}

The CoRoT-3 lightcurve clearly contained a long-term variation, as can
be seen in Fig.~1, where we plot the {\it relative} red-channel flux, 
after subtracting and dividing the original flux by its median.

To remove this trend we used a discrete cosine
transform \citep{ahmed74},  
adopted to the unevenly spaced data we had in hand.
We fitted the data with a linear combination of the first $N$ low-frequency cosine
functions

\begin{equation}
\left\{f_i(t_j)=\cos(\frac{2\pi}{2T}i\times t_j); \ i=0,N\right\} \
,{\rm where} \ \ N=Round\left(\frac{2T}{4P_{orb}}\right)=18 \ ,
\end{equation}
$T=152.012\, d$ is the timespan of the observations, 
$P_{orb}=4.2568\, d$ is the
orbital period of CoRoT-3, and $t_j$ is the timing of the $j$-th
measurement. The fitting finds the linear coefficient $a_i$
for each of the cosine functions, so that the fitted model is

\begin{equation}
\mathcal{M}(t_j)=\sum a_i f_i(t_j) \ .
\end{equation}
We then subtracted the model $\mathcal{M}(t_j)$ from the lightcurve.

The general idea was to perform a high-pass filter, so we removed all the
low-frequency cosine components of the lightcurve 
without altering the periodic
modulation of the orbital period.
In a similar manner we also removed the satellite and earth
frequencies, which appeared in the N2 data \citep[e.g.,][]{mazeh09}.

Fig.~1 shows the 
red-channel
lightcurve before and after the removal
of the long-term trend and the satellite and earth modulation. 
The RMS of the cleaned lightcurve
is 904 ppm (parts per million). 
A similar analysis of the blue- and green-channel data yielded 
lightcurves with highly correlated noise, of an RMS of 1600 and 2000 ppm, 
respectively.  
The white-channel lightcurve, which includes the blue and the green data,
is affected by similar problems.
These results supported our decision 
to consider the red-channel data only.

\begin{figure*}
\centering
\resizebox{10cm}{8cm}
{\includegraphics{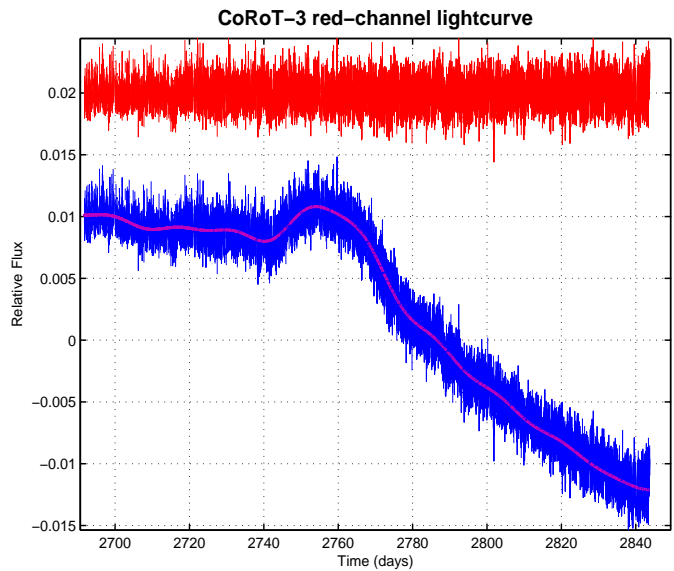}}
\caption{The lightcurve of CoRoT-3, before (blue) and after (red) detrending. The long-term model is presented by the wide line. Time is the CoRoT time = BJD - 2451545.0. Flux is relative to the mean red-channel flux. The detrended lightcurve is shifted by 0.02.
}
\label{detrend}
\end{figure*}

\subsection{Fitting the amplitudes of the ellipsoidal, beaming, and
reflection effects}
We finally proceeded to fit a model that includes the ellipsoidal,
beaming and reflection effects. We approximated each of the three
effects using pure sine/cosine functions, relative to the middle of the
transit, $t_{tran}$, denoted as
phase zero. The reflection and the beaming effects were approximated by
sine and cosine functions, respectively, with the orbital period, and
the ellipsoidal effect by a cosine function with half the orbital
period (see next section). In this approximation, we expressed the stellar
flux modulation
$\Delta F$ as a fraction of the averaged flux ${\bar{F}}$, 
and a function of $\hat{t}\equiv t-t_{tran}$:

\begin{equation} 
\frac{\Delta F_{\mathrm{ellip}}(\hat{t})}   {\bar{F}} =
-A_{ellip}\cos\left(\frac{2\pi}{P_{orb}/2}\hat{t}\right) \ ,
\end{equation} 
  
\begin{equation} 
\frac{\Delta F_{\mathrm{beam}}(\hat{t})}            {\bar{F}}  =
A_{beam}\sin\left(\frac{2\pi}{P_{orb}}\hat{t}\right) \ ,
\end{equation} 
  
\begin{equation} 
\frac{\Delta F_{\mathrm{refl}}(\hat{t})}{\bar{F}} =
-A_{refl}\cos\left(\frac{2\pi}{P_{orb}}\hat{t}\right) \ ,
\end{equation} 
where the coefficients, $A_{ellip}$, $A_{beaming}$, and $A_{refl}$ are
all positive.

We therefore fitted the cleaned, detrended lightcurve of CoRoT-3 with a
5-parameter model, $\mathcal{M}_{ebr}$, consisting of two frequencies

\begin{equation}
\mathcal{M}_{ebr}(t_j)=a_0
+a_{1c}\cos\left(\frac{2\pi}{P_{orb}}\hat{t}_j\right)
+a_{1s}\sin\left(\frac{2\pi}{P_{orb}}\hat{t}_j\right)
+a_{2c} \cos\left(\frac{2\pi}{P_{orb}/2}\hat{t}_j\right)
+a_{2s}\sin\left(\frac{2\pi}{P_{orb}/2}\hat{t}_j\right) \ ,
\end{equation}
as performed by \citet{sirko03}.
The fitting process could find
 any value, positive or negative, 
for the five parameters.
However, we did expect $a_{1s}$ to represent the beaming effect and
therefore be positive, $a_{1c}$ to represent the reflection effect and
therefore be negative, $a_{2c}$ to represent the ellipsoidal effect and
therefore be negative, and $a_{2s}$ to be close to zero. 
The last expectation is important because it ensures that the ellipsoidal modulation
is detected with the correct phase. 
The $a_0$ parameter was induced to remove any DC component left in the data.
Obviously, the
expected absolute values of the three coefficients depended on the
parameters of the CoRoT-3 system. Thus, {\it after} the analysis was performed
we were able to verify our results by comparing them with the expected values.
 
\begin{table}
\caption{The fitted coefficients and the theoretical 
expected amplitudes of the three effects of CoRoT-3}

\begin{tabular}{lrcl}
Coefficient & Derived value   & Expected amplitude & Effect \\
               &   (ppm) \ \ \    &       (ppm)     \\
\hline
$a_{1c}  $ &$-14\pm 9$    &   $-\alpha_{refl}(7.2\pm0.3)$    &
Reflection \\
$a_{1s}  $ &$ 27\pm 9$    &   $\alpha_{beam}(29\pm0.5)$   & 
Beaming \\
$a_{2c}  $ &$-59\pm 9$   &    $-\alpha_{ellip}(32\pm5)$       & 
 Ellipsoidal \\
$a_{2s}$ &$ 0.1\pm 9$ & --- & --- \\
\hline
\end{tabular}
\label{table_coeff}
\end{table}
   
\begin{figure*}
\centering
\resizebox{10cm}{8cm}
{\includegraphics{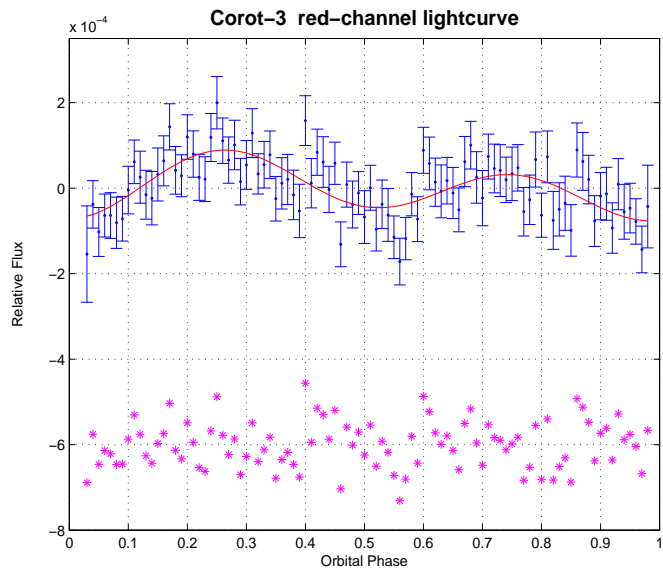}}
\caption{The folded cleaned lightcurve of CoRoT-3, binned into 100
bins,with the fitted model. The residuals are plotted at the bottom of
the
figure}
\label{model_fit}
\end{figure*}

The results of the fitting are given in Table~\ref{table_coeff} 
(with some order-of-magnitude theoretical expectations; see next section) and
plotted in Fig.~\ref{model_fit}. In the figure one can easily discern
the ellipsoidal modulation, with half the orbital period, and the
beaming effect, which causes the difference between the two peaks. 
One indication of the consistency of our results with the expected
modulations is the
correct
sign of the three first coefficients and that the fourth coefficient 
is smaller than the third one by at least one order of magnitude. 
This suggests a close agreement between the orbital phase of the ellipsoidal 
modulation and that of the transit.
We note that the errors in the four coefficients are all 9 ppm, 
or somewhat smaller than $10^{-5}$ in relative flux. This precision does not 
allow a significant detection of the reflection modulation. However, the
detection of both the ellipsoidal and the beaming effect is highly
significant. A bootstrap test indicates that
the probability of detecting the beaming modulation by chance is 
$2 \times 10^{-4}$.

\section{Theoretical approximation}        
\label{theory}                            

This section presents our theoretical approximations
of the ellipsoidal, beaming, and the reflection effects.
We are not interested in detailed calculations,
which depend on specific models. For example,
to calculate the ellipsoidal modulation one could integrate the light 
originating from the individual surface elements of the rotating star, 
which is slightly deformed by the tidal force induced by its small companion
\citep[e.g.,][]{orosz00}.
Such an analysis was carefully performed very recently to model 
the Kepler 
lightcurve of HAT-P-7 \citep{welsh10}. Instead, we are interested here only 
in order-of-magnitude approximation,
so we can check the consistency of the detected amplitudes with
the theory.
All our approximations are evaluated for an inclination angle $i$ close to 90 deg, so we could ignore the $\sin i$ dependence of the three effects.

\subsection{Ellipsoidal effect}

To derive the order-of-magnitude of the ellipsoidal effect we
used
the analytical approximation of \citet{morris93}, 
who used the \citet{kopal59}
expansion of  
the periodic variation into discrete Fourier series with terms that
depend on the ratio $R_*/a$, where $R_*$ is the stellar radius and
$a$ is the semi-major axis of the orbit \citep[see also][]{pfahl08}. 
Assuming $R_*/a$ is
small, the leading term of the stellar variation
has a semi-amplitude of

\begin{equation} 
 A_{ellip}
  \simeq
\alpha_{ellip}
\frac{m_p}{M_*}
\left(\frac{R_*}{a}\right)^3
\ ,
\end{equation} 
where
\begin{equation} 
 \alpha_{ellip}
=
0.15\frac{(15+u)(1+g)}{3-u} \ 
\end{equation}
is of order unity. 
In the above expression, $m_p$ is the planetary mass, 
$M_*$ is the stellar mass,
$g$ is the stellar gravity darkening coefficient,
 and $u$ is its limb-darkening coefficient \citep[e.g.,][]{mazeh08}.  

\subsection{Beaming effect}
For a circular orbit, the amplitude of the beaming effect can be written
as

\begin{equation}
A_{beam}
=
\alpha_{beam}\, 4\frac{K_r}{c} \ ,
\end{equation}
where $K_r$ is the stellar radial-velocity amplitude and $c$ is the
speed of light \citep{loeb03,zucker07}, and $\alpha_{beam}$ is of order
unity. The factor
$4K_r/c$ represents the beaming effect for bolometric photometric
observations, but
ignores the Doppler shift photometric effect, which appears when the
photometric observations are made in a specific bandpass, so that some
of the stellar light is shifted out of or into the observed bandpass. The
latter is accounted for by the $\alpha_{beam}$ factor, and we assume that for
the CoRoT red bandpass it is of order unity.

\subsection{Reflection effect}
In our simplistic approximation we include in the reflection modulation 
the thermal emission from the dayside of CoRoT-3b, 
assuming both are modulated with the same phase
\citep[e.g.,][]{snellen09}. 
The amplitude of the modulation of the reflected light alone is 

\begin{equation}
A_{refl}
=
p_{geo}
\left(\frac{r_p}{a} \right)^2 \ ,
\end{equation}
where $r_p$ is the planetary radius and $p_{geo}$ is the geometrical
albedo \citep[e.g.,][]{rowe08}. 
\citet{rowe08} found quite a small albedo, of 0.03, for HD~209458, but recent study
 \citep{cowan10} suggested that exoplanets may have a much larger albedo, of up to 0.5. We  therefore write the amplitude of the reflection effect, including the thermal emission, as
\begin{equation}
A_{refl}
=
\alpha_{refl}\, 0.1
\left(\frac{r_p}{a} \right)^2 \ ,
\end{equation}
 where $\alpha_{refl}$ is of order unity.

\subsection{Application to CoRoT-3}

Table~\ref{table_corot3} presents the relevant parameters of CoRoT-3. The
first four parameters were derived by \citet{deleuil08}, while the last
parameter, the radial-velocity amplitude, was deduced by
\citet{triaud09}. From these parameters we derived the expected values of
the amplitudes of the ellipsoidal, beaming, and reflection effects, which are given in
Table~\ref{table_coeff}. 

We emphasize that in each of the three theoretical amplitudes given in Table~\ref{table_coeff}, the main source of 
uncertainty is hidden in the $\alpha$ factor, which we did {\it not} calculate.
The numerical values, with their relatively small errors, are only order-of-magnitude approximations
for inclination angles close to 90 deg.
%

\begin{table}
\caption{CoRoT-3 parameters, as derived by \citet{deleuil08} and \citet{triaud09}}
\begin{tabular}{lll}
Parameter & Derived value   & Unit  \\
\hline
$a/R_*$      & $7.8\pm0.4               $    \\
$r_p/R_*$   & $(663\pm9)\times10^{-4}$   &     \\
$m_p$        & $21.7\pm1.0$                      &  $M_{Jup}$   \\
$M_*$        & $ 1.37\pm0.09 $                   &  $M_\odot$    \\
$T_*$         & $6740\pm140 \ K$                  &   deg             \\   
$K_r$          &  $ 2170\pm30 $                    &  m s$^{-1}$  \\
\hline
\end{tabular}
\label{table_corot3}
\end{table}

The amplitudes derived from the cleaned lightcurve of CoRoT-3, as shown
in Table~\ref{table_coeff}, are of the same order of magnitude as the
expected values, based on our simplistic approximation. This is true in
particular for the beaming effect, where the theoretical approximation was
found to be quite accurate. We therefore propose that we have
detected the ellipsoidal and beaming effects of CoRoT-3. 
Our results suggest that the $\alpha_{ellip}$ factor in CoRoT-3 is on
the order of $2$.
The reflection effect was too small to ensure a significant detection,
 given the SNR of the lightcurve.

\section{Discussion}        
\label{discussion}                            
%

Our analysis has demonstrated that the 
red-channel
lightcurve of CoRoT-3 includes the
ellipsoidal and beaming effects. This is the first time that the beaming
effect has been detected for substellar companion. 
We have been able to detect the two effects, with 59 and 27 ppm amplitudes,
respectively, because of a combination of three features:
\begin{itemize}
\item 
CoRoT-3 brightness: With $r'$-mag of 13.1, the star is among the
brightest CoRoT targets, which are typically in the range of 11 to 16
in $r'$ \citep{deleuil08}.
\item
The long observational run: The LRc01 lasted for 152 days, and the CCDs
did not show yet any aging signals.
\item
The mass of CoRoT-3b: This massive-planet/brown-dwarf
companion has
one of the largest masses, 22 $M_{Jup}$, discovered by CoRoT for substellar objects.
\end{itemize}

The last feature
suggests that the stellar rotation has achieved synchronization
with the orbital period of 4.3 days, without which the analysis of the 
 ellipsoidal effect could have been more complicated. 

Had this analysis been performed immediately after the
discovery of the transits of CoRoT-3, and in particular before the
radial-velocity confirmation of the planetary nature of the transiting
object, we could have estimated the mass of the unseen object from the
observed amplitudes of the ellipsoidal and beaming effects, 
provided we had been able to accurately
derive their expected values. Such analysis could, in
principle, save costly radial-velocity observations, or at least
reduce to a minimum the number of observed velocities needed to confirm
the substellar mass of the transiting object.

Obviously, the analysis of lightcurves obtained by space missions is dramatically
 different from those obtained by ground-based photometry. 
For the latter, the appearance of the ellipsoidal modulation
in the data of 
transit candidates 
was
considered a sign that the transiting object 
was 
of stellar nature, as suggested by \citet{sirko03}, and applied, for example, by \citet{kane08} and  \citet{pietrukowicz10}. This is so because of the relatively high threshold of detection of the ellipsoidal modulation in the ground-based photometry.
On the other hand, the present work, and the study of \citet{welsh10}, 
suggest that the detection of the ellipsoidal modulation with 
a small amplitude in the CoRoT and Kepler data 
may indicate that the transiting object is a massive-planet/brown-dwarf object. 
      
The present analysis suggests that, in principle, the three effects, or at
least two of them, can be
detected in the CoRoT lightcurves for
some massive-planet/brown-dwarf objects, even without any transits, as
suggested by \citet{loeb03} and \citet{zucker07}. The effects can be
stronger for systems with shorter orbital periods, and therefore can
be detected in stars fainter than CoRoT-3 in the CoRoT fields.
Many objects similar to CoRoT-3 should also be detected by Kepler, because of both
higher SNR and longer timespan of its lightcurves.


\begin{acknowledgements}
We thank Shay Zucker for helpful discussions, Antonio Claret and Edward Guinan for help with the theoretical estimation of the ellipsoidal modulation, and Avi Shporer for careful reading of the paper and helpful suggestions.  
This research was supported by the ISRAEL SCIENCE FOUNDATION (grant No.
655/07).
\end{acknowledgements}


{}
\end{document}